\documentclass[a4paper]{jpconf}
\usepackage{graphicx}
\usepackage{nicefrac}
\usepackage{enumitem}
\usepackage{amsmath,amssymb,mathrsfs}

\begin{document}
\title{Fractional statistics of charge carriers in the one- and two-dimensional t-J model. A hint for the cuprates?}

\author{P. A. Marchetti$^*$}

\address{Dipartimento di Fisica e Astronomia, Universit\'a di Padova, INFN,\\
Padova, I-35131,Italy\\
$^*$E-mail: marchetti@pd.infn.it}

\begin{abstract}
We show that we can interpret the exact solution of the one-dimensional t-J model in the limit of small J in terms of charge carriers with both exchange (braid) and exclusion (Haldane) statistics with parameter 1/2. We discuss an implementation of the same statistics in the two-dimensional t-J model, emphasizing similarities and differences with respect to one dimension. 
In both cases the exclusion statistics is a consequence of the no-double occupation constraint.
We  argue that the application of this formalism to hole-doped high Tc cuprates and the derived  composite nature of the hole give a hint to grasp many unusual properties of these materials.
\end{abstract}
\section{Introduction}
This paper is a brief review of the attempt to assign 1/2 Haldane statistics to the charge carriers of the one- and two-dimensional $t$-$J$ model, comparing the two cases and arguing that the second case is relevant for the low-energy physics of hole-doped high $T_c$ cuprates. The emphasis is on ideas and only few mathematical details are given.

In quantum systems of identical particles one can define two kinds of statistics. One is the exchange (or braid) statistics:  the ($\bf C$-valued) many-body wave-function acquires a phase factor $e^{i(1-\alpha)\pi}$, with $\alpha \in [0,2[$, when one performs a positively oriented exchange between two particles and the inverse phase factor for a negatively oriented exchange. The parameter $\alpha$ is 0 for fermions, 1 for bosons, and for the other values the corresponding particles are generically called anyons (see e.g. \cite{Wilczek}), in particular for $1-\alpha=\pm 1/2$ they are called semions and in this paper we will be mainly interested in this case. The same phase factors arise in the quantum field theory setting if we perform an equal-time oriented exchange of the fields creating the corresponding particles \cite{FM89}. Anyons require an hard-core exclusion condition and can exist only in space-dimension $D<3$,  where the exchange of two identical hardcore particles can be oriented, namely, one can distinguish an exchange and its inverse. This is possible in one dimension (1D) because the real line is oriented and in two dimensions (2D) because one cannot continuously connect exchanges with opposite orientation. Since oriented exchanges generate the braid groups \cite{Artin}, this statistics is also called braid statistics. This connection can be made more intuitive using the path-integral formalism, where, due to the boundary conditions in time, each configuration of the worldlines of the particles form a braid if there are no intersections. (Actually, using the fact that the probability for two brownian paths in $\bf R^2$ to intersect each other at a fixed time is zero, the hard-core condition is not necessary in 2D in the continuum \cite{FKM92})

 Another kind of statistics can be defined for quasi-particles in finite-density quantum systems of identical particles:  the Haldane's fractional exclusion statistics, generalizing Pauli exclusion principle for fermions \cite{Haldane}. It is  identified by a parameter $g$ measuring the effective interaction among quasi-particles occupying the same state in the one-particle Hilbert space. A way of characterizing this statistics at $T=0$ is the following \cite{Wu}:  A quasi-particle obeys exclusion statistics with parameter $g$ if the quasi-particle density, denoted by $n_g$, having the same volume enclosed by the Fermi surface of the Fermi gas with density $n_0$, satisfies 
 \begin{equation}
\label{eq0}
n_0/n_g = (1 - g).
\end{equation}
Clearly, $g = 0$ for standard fermions and other intermediate values of $g$ in $(0,1)$  define a fractional exclusion statistics. In particular, and this will be the case of interest here, this implies that at a fixed momentum (neglecting other internal degrees of freedom) a quasi-particle with exclusion statistics 1/2 can have an occupation number twice that of a free fermion, so that the volume of its Fermi surface is half of that of a Fermi gas with the same density.

In this paper, following \cite{MSY96, Weng97, PRB19}, we show that we can attribute consistently both exchange and exclusion statistics 1/2 to the charge carriers of the one- and two- dimensional $t$-$J$ model. A quite interesting physical application of these ideas is a derived explanation of many features of the low-energy physics of high $T_c$ cuprates. In fact, most of the researchers believe that for the hole-doped cuprates the key actors in the superconducting transition are the so-called Zhang-Rice singlets \cite{Zhang}, that can be modeled as empty sites of a 2D $t$-$J$ model on a square lattice representing the Cu sites in the CuO planes of the cuprates.  
 The model hamiltonian for a plane is then given by:
\begin{equation}
\label{eq1}
H_{t-J}=\sum_{\langle i, j \rangle} P_G \left[ -t c^*_{i \alpha} c_{j \alpha} + J {\vec S}_i \cdot {\vec S}_j \right] P_G,
\end{equation}
 where $i,j$ denote nearest neighbour (nn) sites of the lattice, $c$ the hole field operator, $P_G$  the Gutzwiller projection eliminating double occupation, $\alpha$ the spin indices and  summation over repeated spin (and vector) indices is understood, here and in the following. For the cuprates a typical value for the nn hopping is $t \approx 0.3 \hspace{2pt} \mathrm{eV}$ and for the anti-ferromagnetic Heisenberg coupling is  $J \approx 0.1 \hspace{2pt} \mathrm{eV}$. (Actually to get a reasonable shape of the Fermi surface in the tight binding approximation one adds to the nn hopping term at least a nnn term with coefficient $t'$, $|t'| \sim 0.03 - 0.1 \hspace{2pt} \mathrm{eV}$, strongly material dependent, but it will not be considered here, except in the final section). As already mentioned we will be interested also in the 1D version of the model which can be exactly solved by Bethe ansatz \cite{Ogata} and conformal field theory techniques \cite{Ren}, but, following \cite{MSY96, Weng97}, here will be solved via spin-charge decomposition. This solution suggests the main strategy to treat the physically more relevant 2D version.
 
Probably the most interesting features of this approach applied to the cuprates are the following: First, the  holes are composites made of only weakly bound spinless holons, charge carriers with Fermi surface,  and spinons, spin carriers without Fermi surface, so that some physical responses  dominated by the spinons have a totally non-Fermi liquid character. This occurs in spite of the fact that the holes have, in the overdoped region, a Fermi surface satisfying Luttinger theorem, precisely due to the 1/2 Haldane statistics of the holons. Second, the semionic statistics of the holons, generated by attached charge vortices, implies the appearance, in the spin counterpart, of  antiferromagnetic spin vortices with opposite chirality when centered on holons in two different N\'{e}el sublattices. Lowering the temperature such gas of vortices undergoes a Kosterlitz-Thouless-like transition, with the formation of a finite density of vortex-antivortex pairs and since the vortices are centered on the charge carriers, this provides a novel topological mechanism of charge-pairing finally leading to superconductivity \cite{PRB11}.

 %%%%%%%%%%%%%%%%%%%%%%%%%%%%%%%%%%%%%%%
\section{The spin-charge decomposition}
Let us outline the formalism of spin-charge decomposition that we use. Inspired by an idea pioneered by Anderson \cite{BZA} and Kivelson \cite{kiv}, suggested for the cuprates by the rather different response of charge and spin degrees of freedom in many experiments, one can tackle the no-double occupation constraint enforced by the Gutzwiller projector $P_G$ by rewriting the hole field $ c_{\alpha}$ as a product of a charged spinless fermion field $h$, the holon, and a neutral spin 1/2 boson field $\tilde s_\alpha$, the spinon, as $c_\alpha=h^* \tilde s_\alpha$ \cite{msy98}. (We use the tilde here because we will denote by $s_\alpha$ a slightly different field).  Being spinless, by the Pauli principle the holon $h$ implements exactly the Gutzwiller constraint. Furthermore if one imposes the constraint $ \tilde s^*_\alpha \tilde s_\alpha=1$, then, since  $ c^*_\alpha c_\alpha=1-h^* h$, one sees that $h^* h$ is just the density of empty sites in the model. However, if we treat the holon in mean-field, because it is spinless the Fermi surface that we get is not the desired one. In 1D , from the exact solution, and in 2D, from experiments for overdoped cuprates (see e.g. \cite{Hus}) , we know that the Fermi surface satisfies the Luttinger theorem, hence the D-volume enclosed by the Fermi surface of fermionic holons is twice the desired one. This suggests that an exclusion statistics with parameter 1/2 would be the appropriate solution to this problem. We now prove that precisely this exclusion statistics appears if the holons are not fermions but semions and show that this exchange statistics can be supported by an argument of energy optimization. To implement the semionic exchange statistics in the path-integral formalism we start from 2D and make use of the following

{\bf  Theorem}
\cite{fro92}\cite{MSY96}
We embed the lattice of the 2D $t$-$J$ model in a 2-dimensional space and denote by $x=(x^0,x^1,x^2)$ the coordinates of the corresponding 2+1 euclidean space-time, $x^0$ being the euclidean time. We couple the fermions of the $t$-$J$ model to a $U(1)$ gauge field,
$B_\mu$, gauging the global charge symmetry, and to an $SU(2)$ gauge field,
$V_\mu$, gauging the global spin symmetry of the model, and we assume that
the dynamics of the gauge fields is described by the Chern-Simons action $-2  S_{c.s.}^{U(1)} (B) + S_{c.s.}^{SU(2)} (V)$, with
\begin{displaymath}
   S_{c.s.}^{U(1)} (B) = \frac{1} {4 \pi} \int d^3 x \epsilon_{\mu\nu\rho}
   B_\mu \partial_\nu B_\rho (x),
\end{displaymath}
\begin{equation}
 S_{c.s.}^{SU(2)} (V) = \frac{1} {4\pi} \int d^3 x {\rm Tr} \epsilon_{\mu\nu\rho}
   [V_\mu \partial_\nu V_\rho + \frac{2}{3} V_\mu V_\nu V_\rho](x),
\label{eqCS}   
\end{equation}
where $\epsilon_{\mu\nu\rho}$ is the Levi-Civita anti-symmetric tensor in 3D and $\partial_\nu=\partial/\partial x^\nu$.
Then the spin-charge (or $SU(2) \times U(1)$) gauged model so
obtained is exactly equivalent to the original $t$-$J$ model.
In particular the spin and charge invariant euclidean correlation functions of the fermions fields  $c_{j \alpha}$ of the $t$-$J$ model are exactly equal to the correlation functions of the fields $ \exp[i \int_{\gamma_j} B] P(\exp[i \int_{\gamma_j} V])_{\alpha \beta} c_{j \beta}$ , where $c$ denotes now the fermion field of the gauged model, ${\gamma_j}$ a string at constant euclidean time connecting the point $j$ to infinity and $P(\cdot)$ the path-ordering, which amounts to the usual time ordering $T(\cdot)$, when ``time'' is used to parametrize the curve along which one integrates.

The two crucial points in the proof of the above theorem are: First, due to a result of Witten \cite{Witten} the integration of the two above gauge fields on a set of closed string forming a braid averaged with the above Chern-Simons actions exactly cancel against each other. Second, the worldlines of the fermions of the $t$-$J$ model in the partition function (joined to the strings in the correlation functions) form braids , because they have no intersections, due to the Gutzwiller constraint.

To the fermion field of the gauged model we now apply the spin-charge decomposition discussed above, rewriting it as a product of a holon and a spinon field. A good feature
of introducing the above gauge fields is that they allow
a more flexible treatment of charge and spin responses
within a spin-charge decomposition scheme, attaching the "charge" $B$-string to the holon and the "spin" $V$-string to the spinon. Spin and charge responses are in fact  quite independent in 1D and this turns out to be partially true even in 2D. One can verify \cite{fro92,MSY96} that the non-local quantum field operator corresponding to the product of the fermionic holon field $h$ and the "charge gauge string" $ \exp[i \int_{\gamma_j} B]$ obeys semionic statistics and the same is true for the product of the bosonic spinon field $\tilde s$ and the "spin gauge string" $P \exp[i \int_{\gamma_j} V]$.
After the holon-spinon decomposition introduced above has been implemented the action of the gauged $t$-$J$ model at doping $\delta$ is given by \cite{msy98} :
\begin{eqnarray}
\label{eq2} 
&S(h,h^*, \tilde s^*, \tilde s, B,V)=\int_0^\beta dx^0 \sum_j h^*_j [\partial_0 -i B_0(j)-2 t \delta] h_j + i B_0(j)+i(1-\\& h^*_j h_j)( \tilde s_{j \alpha}^* V_0(j)_{\alpha \delta} \tilde s_{j \delta} ) - \sum_{\langle i,j \rangle} t h^*_j 
 \exp[i \int_{\langle i,j \rangle} B]h_i (\tilde s_{i \alpha}^* P(\exp[i \int_{\langle i,j \rangle}  V])_{\alpha \delta} \tilde s_{j \delta})]\nonumber\\& +\frac{J}{2}(1-h^*_jh_j)(1-h^*_ih_i) [|(\tilde s_{i \alpha}^* P(\exp[i \int_{\langle i,j \rangle}  V])_{\alpha \delta} \tilde s_{j \delta})|^2-\frac{1}{2} ]-2  S_{c.s.}^{U(1)} (B) + S_{c.s.}^{SU(2)} (V),\nonumber
\end{eqnarray}
where $\beta=1/k_B T$.
The two-point euclidean correlation function of the hole can be written as:
\begin{eqnarray}
\label{eq3}
& G(x,y) =\langle c^*_{x\alpha}(x^0) c_{y\alpha}(y^0) \rangle =\int \mathcal{D}h\mathcal{D}h^*\mathcal{D}\tilde s\mathcal{D}\tilde s^*\mathcal{D}B \mathcal{D}V  \exp[-S(h,h^*,\tilde s, \tilde s^*,B,V)] \nonumber\\ & h_x \exp[-i \int_{\gamma_x} B](x^0) \exp[i \int_{\gamma_y} B]h^*_y(y^0)(\tilde s_{x \beta}^*P(\exp[-i \int_{\gamma_x} V])_{\beta \alpha})(x^0)(P(\exp[i \int_{\gamma_y} V])_{\alpha \delta}\tilde s_{y \delta})(y^0) \nonumber\\
&[ \int \mathcal{D}h\mathcal{D}h^*\mathcal{D}\tilde s\mathcal{D}\tilde s^*\mathcal{D}B \mathcal{D}V  \exp[-S(h,h^*,\tilde s, \tilde s^*,B,V)]]^{-1},
\end{eqnarray} 
where $x,y$ are lattice sites and  $x^0,y^0$ are euclidean times, which in the following will be often understood, if no confusion can arise.

\section{The one-dimensional $t$-$J$ model}

To discuss the one-dimensional model we perform a dimensional reduction of the gauged model defined in Theorem 1,  by restricting the spatial support of the hole fields to a line, let's say along $x^1$, keeping the strings $\gamma$ not along this line to avoid intersections with the hole worldlines making the previous theorem inapplicable.

In 1D one can give an intuitive picture of holons and spinons as follow (see e.g.\cite{Gia}): Consider the Heisenberg spin 1/2 chain, describing the 1D $t$-$J$ model in the limit of zero doping and take as reference state the one with the spin antiferromagnetically ordered,  mimicking the N\'{e}el order appearing in 2D. If we insert a dopant by removing the spin from a site then the two neighbouring spins will be ferromagnetically aligned. Let then the empty site to hop by a simultaneous opposite hopping of the spin, then we get two separate excitations. There is an empty site, but with neighbouring spins antiferromagnetically aligned, thus carrying charge but not spin; the corresponding excitation is the holon. There is another site where one finds a domain wall between two different N\'{e}el sublattices, hence carrying spin 1/2 but neutral; the corresponding excitation is the spinon. Notice, however, that attached to the site with a spin mismatch of 1/2 there is a string of spins flipped w.r.t. the reference state from that site to the holon position, which is integral part of the spinon excitation and corresponds to the "spin string" introduced in Theorem 1.
In the continuum semiclassical limit one can view the spinon as a spin kink, continuum analog of the above domain  wall.

Let's make the above picture mathematically precise in the spin-charge decomposition formalism previously outlined. Both in 1 and 2 D the gauged model equivalent to the $t$-$J$ model has three local gauge invariances: the "charge" $U(1)$, the "spin" $SU(2)$ and the slave-particle corresponding to the multiplication of the holon field $h$ and the spinon field $\tilde s_\alpha$ by the same local phase, thus leaving the hole field unchanged.

In 1D \cite{MSY96} we gauge-fix the first symmetry by setting $B_2=0$, the second one, as suggested by the previous "picture", by setting $\tilde s_{j \alpha} = \sigma_x^j (1 0)^t$, where $\sigma_x$ denotes the Pauli matrix and the superscript $t$ the transpose. Finally the slave-particle gauge symmetry is fixed by imposing a kind of 1D Coulomb gauge: $\arg(\tilde s_{\alpha i}^* P(\exp[i \int_{\langle i, j \rangle} V])_{\alpha \beta}\tilde s_{\beta j})=0$. We now turn to the discussion of charge and spin carriers. We start with the "charge carrier".

Since the 0-component of $B$ appears linearly in (\ref{eq2}) , we can safely integrate it out getting the constraint (with $\mu,\nu=1,2, z \in \mathbf{R}^3$):
\begin{eqnarray}
\label{eq4}
\epsilon_{\mu\nu}\partial_\mu B_\nu(z)=- \pi \delta (z^2) [\sum_j \delta(z^1-j)(1-h^*_jh_j)(z^0)].
\end{eqnarray}
By imposing the gauge-fixing $B_2=0$ one finally gets
\begin{eqnarray}
\label{eqB}
B_1(z)=-\frac{\pi}{2} {\rm sgn}(z^2) [\sum_j \delta(z^1-j) (1-h^*_jh_j)(z^0))],
\end{eqnarray}
with the convention $ {\rm sgn}(0)=0$, so that the holon field turns out to be given by
\begin{eqnarray}
\label{eq5}
  \exp[i \int_{\gamma_j} B]h_j=\exp[\pm i \frac{\pi}{2}\sum_{\ell>j} (1-h^*_\ell h_\ell)] h_j, 
\end{eqnarray}
with the sign in the exponent being positive (negative) if the angle between the $x^1$ axis and $\gamma$ is positive (negative) in $]-\pi, \pi[$. Therefore the sign depends on the choice of the orientation of the $x^1$ line in the spatial plane; since the original model is independent of this choice we take an average of the two possibilities. If the doping is $\delta$, the fermion $c_\alpha$ of the 1D $t$-$J$ model in the tight-binding approximation has a Fermi momentum $ \pi(1-\delta)/2$ since two fermions with opposite spin can have the same momentum. For the spinless fermion $h^*$ the Fermi momentum would be $ \pi(1-\delta) $ since only one spinless fermion con have a fixed momentum. However, if we consider the phase  string attached to the spinless fermion in (\ref{eq5}) we see that, since the expectation value of $h^*_\ell h_\ell$ is $\delta$, in the leading term it contributes to the Fermi momentum a term $-\pi(1-\delta)/2$. Hence the distance between the two Fermi point of the semionic holon of (\ref{eq5}) is $\pi(1-\delta)$ and it obeys an Haldane statistics of parameter 1/2. 

To derive the holon contribution to the euclidean $c$-correlator $G(x,y)$ in the scaling limit  we
decompose each $h$ into right and left movers as
$h(x) \sim h_{R}(x)e^{i \pi (1-\delta) x^1} + h_{L}(x) e^{-i \pi (1-\delta) x^1}$ , the spinor doublet
$H=(h_{R},h_{L})^t$ having the
dynamics of a massless Dirac fermion . Using as in \cite{ye} the Schwinger formula \cite{schw} for the two-point correlation functions of massless 2D Dirac fields coupled to gauge fields,  one finds as leading holon contributions corresponding to the two string directions: $e^{\pm i \pi (1-\delta) x^1/2 } [(x^1-y^1) \pm i v_c (x^0-y^0)]^{-1/2} [(x^1-y^1)^2 + v^2_c (x^0-y^0)^2]^{-1/16},$ where $v_c \sim t_R$ denotes the charge velocity.

Let us turn to the "spin carrier".
Since we have already gauge-fixed the $SU(2)$ gauge-invariance using the spinons, the gauge field $V$ has to be integrated without gauge-fixing. This can be implemented by splitting the integration over $V$ into
an integration over a field $V^m$, satisfying the gauge-fixing condition $V^m_2=0$
and its gauge transformations expressed in terms of an
$SU(2)$-valued scalar field $g$,
i.e., $V_\mu= g^\dagger V^m_\mu g+g^\dagger\partial_\mu g, \mu=0,1,2$. 
%Let's recall  that $ P(\exp[i \int_x^y V])=g^\dagger_y P(\exp[i \int_x^y \bar V])g_x$.
We now find the configuration of $g$ that optimize the partition function of holons in a fixed, but holon-dependent, $g$-background.
It is rigorously proved in Ref.\cite{MSY96} that such configuration is given by $g^m_j=\exp[ -i \frac{\pi}{2} \sigma_x \sum_{\ell>j} h^*_\ell h_\ell ]$. This is the ``string'' of spin flips that we have encountered in the previous "pictorial" discussion. 
Finally we set $g=Ug^m$, with $U \in SU(2)$ describing the fluctuations around the optimal configuration. We write 

\begin{eqnarray} 
\label{eq6}
U= \begin{pmatrix}
 s_1 &-s_2^*\\ s_2 & s_1^*
\end{pmatrix},
\end{eqnarray} 
and we will call in the following the $s_\alpha$ again ``spinons''.
As a first ``mean field'' approximation we
 neglect in the calculation of $V^m$ the fluctuations of the spinons in the directions orthogonal to the spins of the optimal configuration. Proceeding now as done for the $B$ field and integrating over  $V^m_0$ one finds
\begin{eqnarray}
\label{eq7}
V^m_1(z)=\frac{\pi}{2} {\rm sgn}(z^2) [\sum_j \delta(z^1-j) (1-h^*_jh_j)(z^0))(s_{1j}^* s_{1j}-s_{2j}^*s_{j2})]\sigma_3.
\end{eqnarray}
Let us remark an interesting feature \cite{msy07} of the above approach: thanks to the holon-depending spin flips, since the holon in a oriented hopping link is at its end,  $(\tilde s_{i \alpha}^* P(\exp[i \int_{\langle i,j \rangle}  V]))_{\alpha \delta} \tilde s_{j \delta})$ in the $t$-term of (\ref{eq2}) equals $(\sigma_x^{|i|}U_i^\dagger U_j \sigma_x^{|i|})_{11}=s^*_{\alpha i}s_{\alpha j}$, whereas in the $J$-term, where there are no holons, it equals  $(\sigma_x^{|i|}U_i^\dagger U_j \sigma_x^{|j|})_{11}=\epsilon_{\alpha \beta} s_{\alpha i}s_{\beta j}$. But under the constraint  $s^*_{\alpha j}s_{\alpha j}=1$ the following identity holds: $|s^*_{\alpha i}s_{\alpha j}|^2+|\epsilon_{\alpha \beta} s_{\alpha i}s_{\beta j}|^2=1$ so that if one optimizes the $t$-term choosing $|s^*_{\alpha i}s_{\alpha j}|=1$, simultaneously one optimizes also the $J$-term. This optimizing property is somewhat strange, since the same expression in terms of $V$ gives rise to different expressions in terms of the spinons $s$ in the $t$- and the $J$-terms. It is intrinsically due to the $SU(2)$ gauge degrees of freedom, absent in the more standard slave-particle approaches  involving only $U(1)$ gauge degrees of freedom. An exception is the "string formalism" of Weng \cite{Weng97}, in 1D essentially equivalent to the charge-spin decomposition discussed here. The optimization is achieved there using the Gutzwiller projection in the squeezed chain (see below) of occupied sites, the only ones in which spinons are defined in this Hamiltonian formalism, a feature that however has no natural extension to 2D.

As a consequence of the spin flips the gauge-fixing for the slave-particle gauge symmetry in the present formalism becomes $\arg( s_{\alpha i}^* s_{\alpha j})=0$ in the $t$-term, but in the links of the $J$-term it is given by  $\arg( s_{\alpha i}^* \epsilon_{\alpha \beta} s_{\beta j})= 0$. This fact is used in the following further approximations.
 We assume that the spin fluctuations can be treated in mean field in the hopping term because the corresponding term is real due to the discussed slave-particle gauge fixing  and we denote by $t_R$ the renormalized hopping. Furthermore
since the paths of spinons cannot overlap due to the Gutzwiller constraint we replace for the spinon motion the original chain by a squeezed chain of lattice spacing $(1 - \delta)$ with single occupancy constraint. We redefine the spinons in the squeezed chain, denoted by $\mathrm{s}$, through $\mathrm{s}_{\alpha j}= (U_j \sigma_x^{|j|})_{\alpha 1}$. Written in terms of $\mathrm{s}$ the $J$-term is just the Heisenberg model in the Schwinger-boson representation. The slave-particle gauge fixing becomes $\arg( \mathrm{s}_{\alpha i}^* \mathrm{s}_{\alpha j})=0$. Then  approximating the term quartic in $\mathrm{s}$ by mean field and making using of the reality condition imposed by the gauge-fixing, the $J$-term can be rewritten as
\begin{eqnarray}
\sum_i \mathrm{s}_{\alpha i}^* \partial_\tau \mathrm{s}_{\alpha i} + J_R \sum_{\langle i,j \rangle} ( \mathrm{s}_{\alpha i}^*\mathrm{s}_{\alpha j} + h.c.) + i \sum_i \lambda_i ( \mathrm{s}_{\alpha i}^*  \mathrm{s}_{\alpha i} -1)
\end{eqnarray}
with $\lambda$ a Lagrange multiplier field enforcing the single-occupancy constraint and $J_R$ a renormalized spin coupling constant. (The absence of a possible gap term derived from umklapp is justified by its absence in the Heisenberg model, see e.g. \cite{fradkin}). 

One can now compute the long-wavelength continuum limit of the spinon contribution to the euclidean correlator $G(x,y)$ of the fermion $c$ of the $t$-$J$ model. Motivated by \cite{MSY96}, but following  closer the formalism of \cite{Weng97}, one first fermionize separately the two components of  $\mathrm{s}$ by a Jordan-Wigner transformation and we denote by $f_\alpha$ the corresponding fields.
Reinserting the "spin string", the semionic spinon field in the squeezed chain is written in terms of $f$ as $f_{\alpha j} \exp[(-1)^\alpha (\pm) i \frac{\pi}{2}\sum_{\ell>j} f^*_{\beta \ell} f_{\beta \ell})]$, where the sign $\pm$ depend on the direction of the "spin string" w.r.t. the $x^1$ axis, as discussed for the holon. Similarly, in the scaling limit we
decompose each $f_\alpha$ into right and left movers as
\begin{eqnarray}
\label{eqf}
f_\alpha(x) \sim f_{\alpha R}(x)e^{ik_fx^1} + f_{\alpha L}(x) e^{-ik_fx^1},
\end{eqnarray}
$k_f$ being the spinon Fermi momentum in the squeezed chain, with  massless Dirac  spinor doublets
$F_\alpha=(f_{\alpha R},f_{\alpha L})^t$, plus the single-occupancy constraint. In the leading term the mean-field contribution of the exponential in the "spin string" exactly cancels one of the two  $k_f$ exponentials in (\ref{eqf}) so that in this approach the spinons do not contribute to the Fermi momenta of the fermion $c$ of the $t$-$J$ model. Hence, taking into account the Haldane statistics of the holon, when the electron is reconstructed combining the holon with the spinon, it has the same Fermi surface of the tight-binding approximation for the fermion, as we wanted to prove, thus satisfying Luttinger theorem.

For completeness we now outline how to finalize the computation of the $c$ correlator.
To implement exactly the single-occupancy constraint we use the technique of duality (equivalent to bosonization in 1D \cite{que,fro95, mar95}): we couple minimally each $F_\alpha$ with a $U(1)$ gauge field $A_{\alpha \mu}, \mu = 0,1$ and impose the constraint that the field strengths of $A_{\alpha}$ are zero, by inserting in the Lagrangian a term with multiplier fields, denoted by $\phi_\alpha$, given by  $i \epsilon_{ \mu\nu}\partial_\nu \phi_\alpha A_{\alpha \mu}$. Defining $\phi_\pm = (\phi_1 \pm \phi_2)/2$ a change of variable, re-adsorbing $\lambda$ into the gauge fields $A_{\alpha}$ produces the constraint $\phi_+=0$. Then using the procedure of \cite{ye} integrating out $A_\alpha $ and $\phi_-$ one gets for the spinon contributions with the two directions of the strings: $ [(x^1-y^1) \pm i v_s (x^0-y^0)]^{-1/2} $,  where $v_s \sim J_R$ denotes the spin velocity.
Combining the holon and spinon contribution one thus  reproduces \cite{MSY96, Weng97} the correct long-wavelength limit of the correlation function $G(x,y)$ of the 1D $t$-$J$ previously obtained by Bethe ansatz and conformal field theory techniques .

\section{The two-dimensional $t$-$J$ model}

If we try naively to export to 2D the string mechanism discussed in 1D we immediately find a big difference already in the intuitive picture: in correspondence to the string  of spin flips between the holon and spinon positions in a chain, the two parallel adjacent chains have the spin with the same orientation of those of the string. and since the spin interaction is antiferromagnetic this configuration costs an energy proportional to the length of the string, i.e. holon and spinon are confined: this is a manifestation of the slave-particle gauge force. However, this is not an optimal solution from the point of view of energy. To get a suggestion for improvement one  notices that the kink corresponding to the continuum limit of the spin string is the typical semiclassical excitation in 1D, but not in 2D, where its role is played by the vortex.
If we accept the suggestion coming from the 1D model for the exchange statistics of holon and spinon, one should search for a semionic representation of the hole field also in 2D and precisely a charge-vortex attached to a fermionic "bare" holon converts it into a semion. As explained later, the spin-vortex generated by consistency will also play a key role in the formalism. These vortices are somewhat analogous to those introduced by Laughlin in
the Fractional Quantum Hall Effect and in fact a semionic representation of the hole was advocated by him \cite{La} quite soon after the discovery of high $T_c$.
As we sketch below, in 2D, as in 1D, this semionic holon obeys a 1/2 exclusion statistics.

We now make these ideas precise in  the spin-charge gauge formalism. 
%In 2D we use the Coulomb gauge-fixing $\partial_\mu B^\mu=0$ for the "charge" gauge symmetry,  and the gauge-fixing used in 1D ,  $\tilde s_{j \alpha} = \sigma_x^j (1 0)^t$, for the "spin" gauge symmetry.
 For convenience we start with the spin excitation, since some of its features will be needed in the description of the charge carrier, but we try to follow as close as possible the procedure discussed in 1D. Analogously we introduce the gauge-fixing   $\tilde s_{j \alpha} = \sigma_x^j (1 0)^t$  for the "spin" gauge symmetry, the field $V^m$ satisfying the Coulomb gauge-fixing condition  $\partial_\mu V^m_\mu=0, \mu= 1, 2$ and its gauge transformations, again written in terms of an $SU(2)$-valued $g$ field.
 We didn't succeeded to find rigorously, as in 1D, a configuration of $g$, depending on the holon configuration, optimizing the holon-partition function in that $g$ background, but we still found a configuration $g^m$ optimal ``on average'', in a Born-Oppenheimer approximation \cite{msy98}. As in 1D in this configuration the 
 spinons $\tilde s$ are antiferromagnetically ordered but there  is in addition a spin flip on the sites where holons are
 present, as for the final site of a hopping link of holons at the time of hopping. Hence, neglecting $V^m$,  the argument given in the previous section  on the simultaneous optimization in this formalism of the $t$ and $J$ terms holds also in 2D, giving a motivation ``a priori'' for this spin-charge gauge approach.
 
 Spin fluctuations around the "optimal" configuration are described by the $SU(2)$- matrix field $U$ whose components are written, as in 1D, in terms of spinon fields $s_\alpha$. At least in the long wavelength continuum limit it is convenient to make the slave-particle gauge symmetry explicit by introducing the related gauge field that we denote by $A_\mu$. This can be done in absence of holons simply by assuming a continuum limit of the spinon field $s$ of the form
$s_{x \alpha}(x^0) \longrightarrow s_\alpha(x) + (-1)^{|x|} p_\alpha(x)$, where $s$ and $p$ in the r.h.s. are continuum fields and $|x|=x^1+x^2$. Integrating out the ferromagnetic component $p$, as shown in \cite{sac}, both in 1D and 2D we obtain a low-energy spinon Lagrangian in the form of a CP1 model: 
\begin{eqnarray}
    \mathcal{L}_s=   \frac{1}{g}
   \left[ \left| \left( \partial_{0} - i A_{0} \right) s_{\alpha}
   \right|^{2} + v_{s}^{2} \left| \left( \vec \nabla - i \vec A
   \right) s_{\alpha} \right|^{2} 
   \right],
   \label{eqCP}
\end{eqnarray}
where  $g \sim J$ and $v_s \sim J $ is the spinon velocity, with the lattice spacing set to 1. Furthermore $A_\mu(x)=(-1)^{|x|} s^*_{\alpha}(x)\partial_\mu s_\alpha(x)$, with $\mu$ a space-time index, and the implicit constraint $s^*_{\alpha} s_\alpha=1$ is understood. For later purposes we remark that in 1D there is an additional $\Theta$-term
\begin{eqnarray}
\label{eq8}
    \mathcal{L}_\Theta= i \frac{\Theta}{2 \pi} \epsilon_ {\mu\nu}\partial_\mu A_\nu,
\end{eqnarray}
with $\Theta = \pi$.

In 2D neglecting, with a stronger approximation w.r.t. 1D , the spinon fluctuations $U$ in the computation of $V^m$ one gets ( with $\mu=1,2)$:
\begin{eqnarray}
\label{eq9}
V^m_\mu(z)= -\frac{1}{2} \sum_j (-1)^{|j|} \partial_\mu\arg(\vec z-j) h^*_jh_j(z^0) \sigma_z,
\end{eqnarray}
with $|j|=j^1+j^2$.
We recognize in the term $(-1)^{|j|} \partial_\mu\arg(\vec z-j)$ the vector potential of a vortex centered on the holon position $j$, with opposite vorticity (or chirality) for the center in opposite N\'{e}el sublattices. We call these vortices antiferromagnetic spin vortices; they are the topological excitations of the $U(1)$ subgroup of the $SU(2)$ spin group  unbroken in the antiferromagnetic phase.  Hence  they are still a peculiar manifestation of the antiferromagnetic interaction, like the more standard antiferromagnetic spin waves, but they appear only in 2D. These vortices are of Aharonov-Bohm type, hence a purely quantum effect,  inducing a topological effect far away from the position of the holon itself, where their classically observable field strength is supported. 

In the low-energy continuum limit one can see from (\ref{eq2}), expanding to the leading power the exponentials in the links, that $V^m$ appears linearly in the $t$-term of the action; since its spatial average vanishes in a ``mean field'' treatment we ignore it. With the same approximation in the $J$-term  $V^m$ appears instead quadratically and assuming self-consistently that for sufficiently large $\delta$ the full $SU(2)$ spin symmetry is restored at large scales, one finds an interaction between vortices and spinons proportional to:
\begin{eqnarray}
\label{eq10}
\int d^3x (V^m_\mu V^m_\mu)(x) s^*_\alpha s_\alpha (x).
\end{eqnarray}
A quenched average, $\langle \cdot \rangle$, over the positions of the center of the antiferromagnetic spin-vortices yields the following estimate  \cite{msy98}:  $\langle V^m_\mu V^m_\mu \rangle \approx \delta |\log \delta|$, thus  providing a mass-gap to the spinons. Hence the  gapless spinons $s$ forming the spin waves of the CP1 (or equivalently $O(3)$) model describing the undoped system, traveling in a gas of antiferromagnetic spin vortices centered on holons acquire a gap $ m_s$ , converting the long-range AF of the undoped model in the short-range AF when doping exceeds a critical value. This is selfconsistent with the previous assumption of the $SU(2)$ symmetry restoration at large scales.  Leaving aside for the moment the monomial quartic in the holons of the $J$-term in (\ref{eq2}), we treat in mean field the monomial quadratic in the holons, so that the antiferromagnetic coupling is renormalized to $J (1-2 \delta)$. We see here the effect of strong reduction of antiferromagnetism due to the increase of the density of empty sites, corresponding to Zhang-Rice singlets in the cuprates. As a result one finds that the continuum limit of the term 

\begin{eqnarray}
\label{eq11}
\int dx^0 & \Big[ \sum_j i( \sigma_x^{|j|} U^\dagger_j \partial_0 U_j \sigma_x^{|j|})_{11}+ \sum_{\langle i,j \rangle} \frac{J}{2}(1-h^*_jh_j-h^*_ih_i)\nonumber \\&[ |(\sigma_x^{|i|}  U^\dagger_i  P(\exp[i \int_{\langle i,j \rangle}  V^m]) U_j \sigma_x^{|j|})_{11}|^2 -\frac{1}{2}] \Big](x^0)
\end{eqnarray}
within the above approximations is given by
\begin{eqnarray}
\label{eq12}
\int d^3x \frac{1}{g}
   \left[\left| \left( \partial_{0} - i a_{0} \right) s_{\alpha}
   \right|^{2} +  v_{s}^{2} \left| \left( \vec \nabla - i \vec a
   \right) s_{\alpha} \right|^{2} 
   +m_s^2 s^*_{\alpha}s_{\alpha}
   \right](x),
   \end{eqnarray}
   where $m_s \approx J (1-2 \delta)\sqrt{\delta |\log \delta|}$.
In view of the action (\ref{eq12}) within this formalism it is natural to consider the bosonic spinons $s_\alpha$ as the spin carriers at large scales, at odds with the semionic statistics of the spinons found in 1D also in the continuum limit. However, still a "spin gauge string" will be necessary to reconstruct the hole, as discussed below. The spinon system behaves as a spin liquid
since the spinon confinement which would appear with the massive action (\ref{eq12}) is avoided by  the interaction with the gapless holons, as briefly discussed in the next section. 

We now turn to the charge carrier.
We use the Coulomb gauge-fixing $\partial_\mu B_\mu=0, \mu= 1, 2$ for the "charge" gauge symmetry and integrating the $B_0$ field we obtain $B_\mu(z)=B^m_\mu + b_\mu(z)$ where $B^m_\mu$ introduces a $\pi$-flux phase, i.e. $\exp[\int_{\partial p} B^m]=-1$ for every plaquette $p$ and
\begin{eqnarray}
\label{eq13}
b_\mu(z)=\frac{1}{2}[\sum_j \partial_\mu \arg(\vec z-j)( h^*_jh_j)(z^0))].
\end{eqnarray}
 We can recognize in $ \partial_\mu \arg(\vec z-j)$  the vector  potential of a vortex centered on the holon position $j$ , i.e. centered on an empty site of the $t$-$J$ model. At first we neglect the $b$-field. Then, as discussed in \cite{AM}, through Hofstadter mechanism \cite{hof} the $\pi$-flux (chosen in staggered form) converts the holon   with tight-binding dispersion
$\omega_h \sim 2t[(\cos k_x + \cos k_y) - \delta]$ 
 into a pair of lattice ``Dirac fields'', $\psi_a$, with pseudospin index $a$ related to the two  N\'{e}el sublattices and
with dispersion:
$\omega_h \sim 2t [\sqrt{\cos^2 k_x + \cos^2 k_y }- \delta]$ 
restricted to the magnetic Brillouin zone. One thus obtain two ``small FS'' centred at $(\pm \frac{\pi}{2},\pm \frac{\pi}{2})$
 with
Fermi momenta $k_F \sim \delta$. 

Above a crossover temperature $T^*$ we find that the optimal configuration $g^m$
 involves also a phase factor cancelling the contribution of
$B^m$ in the loops of hopping links of holons, so that the hopping
holons feel an approximately zero flux \cite{msy05} and one recovers the tight-binding Fermi surface. By analogy with the phase diagram of hole-doped cuprates, we call pseudogap (PG) the parameter region in the $\delta$-$T$ plane below $T^*$ and strange metal (SM) the region above.

Reintroducing the $b$-field, the non-local field $h_j \exp[\int_{\gamma_j} b]$ obeys semionic statistics.
We now use a result proven in \cite{ye15}: 

{\bf Theorem} In an incompressible 2D Fermi system with Hall conductivity $\sigma_H= 1/(2 \pi)$, if its fermions are coupled to a Chern-Simons gauge field $b$ with action  $-\frac{1}{\alpha} S_{c.s.}^{U(1)} (b)$ with the statistical parameter $\alpha =1/2$ turning them into semions, then these semions obey the Haldane statistics with parameter 1/2.

Let's sketch the proof. From the Chern-Simons coupling with parameter $\alpha$  we get $\epsilon_{\mu \nu} \partial_\mu b_\nu= -2 \pi \alpha j_0$ with $j_0$ the charge density. Inside a region $S$ we turn on the Chern-Simons coupling $\alpha$  adiabatically from 0 to 1/2. For an incompressible Hall system
the generated statistical flux in turn induces a charge
current perpendicular to the boundary of $S$.
Denoting by $N(\alpha)$ the particle number in $S$ with Chern-Simons coupling $\alpha$ and using the continuity equation $\partial_0 j_0= -\partial_\mu j_\mu, \mu= 1,2$ for the charge current $j_\mu$ and the Hall relation $j_\mu= \sigma_H \epsilon_{ \mu \nu} \partial_0 b_\nu$, we derive
\begin{eqnarray}
\label{eqH}
&\partial_0 N(\alpha) = \int_S d^2x \partial_0 j_0(x) =- \int_S d^2x \partial_\mu \sigma_H \epsilon_{ \mu \nu} \partial_0 b_\nu(x) = \nonumber \\&\int_S d^2x 2 \pi \alpha \sigma_H j_0(x)= 2 \pi \sigma_H \partial_0 \alpha N(\alpha). 
\end{eqnarray}
Integrating over the time of the adiabatic process from (\ref{eqH}) we get $N(1/2)(1- \pi \sigma_H)=N(0)$ and using  $\sigma_H= 1/(2 \pi)$ comparing with (\ref{eq0}) we get $g=1/2$.

Hence, in order to show that the semionic charge carriers of the 2D $t$-$J$ model in the present formalism have an exclusion statistics parameter 1/2 we need to prove that the "bare" fermionic holon system has Hall conductivity  $1/(2 \pi)$ and is incompressible. 
To prove the first property we start noticing that if
the constraint $ \tilde s^*_\alpha \tilde s_\alpha= s^*_\alpha  s_\alpha=1$ is imposed, since $
c^*_\alpha c_\alpha=1-h^* h$, one finds that $(\exp[i \int_{\gamma_j}
B]h_j)^*(\exp[i \int_{\gamma_j} B]h_j)=h_j^* h_j$ is just the
density of empty sites in the model and the Gutzwiller constraint is exactly implemented.
However, in the large-scale continuum limit we have seen that the interaction with antiferromagnetic spin vortices makes the spinon $s(x)$ 
gapped. This implies that in this limit the single-occupancy constraint on $s$
 is not fully satisfied, as the spinon mass gap is
incompatible with it, so the Gutzwiller projection is not anymore
exactly implemented.

To understand the implication let us at first consider the free fermionic holons at doping concentration $\delta=0$  in the presence of the staggered $\pi$-flux implemented by $ B^m_{\langle i,j
\rangle}$. As mentioned above, they are  described by massless
Dirac fields with two flavors,  $\psi_a$, with dispersion given by  two
double-cones with vertices at  $(\pm \pi/2, \pi/2)$ in a suitably defined magnetic Brillouin zone. Since the holons are spinless if they obey fermionic exclusion statistics they fill both the sheets of the double-cones, hence no dopants can be introduced for $\delta \neq 0$. However, if the semionic holons obtained by coupling to the $b$-field satisfy the exclusion statistics with parameter 1/2, at half filling they fill only the lower bands. Still the holon density doesn't vanishes even if there are no empty sites in the original $t$-$J$ model at  $\delta = 0$. The lower bands are thus an artifact produced by the violation of the Gutzwiller constraint   in the large-scale continuum limit. Since the
semionic holons in the lower bands are a result of relaxing the
Gutzwiller projection, they are ``spurious'' and describe the singly
occupied sites in the original unprojected lattice model. When the
doping holes are introduced in the $t$-$J$ model, the corresponding
``physical'' holons partially fill the upper bands of the double-cones and are
responsible for the low energy physics.

Although the spurious lower
bands of holons are not directly relevant to the low energy physics in
the scaling limit, they are responsible for the 1/2 exclusion
statistics. In fact, due to parity anomaly \cite{Re}, to compute the Hall conductivity of
massless Dirac fields we need to introduce an infrared regulator
(like a mass) with a parameter $m$, respecting the symmetry of the
system; at the end of the computation one takes the limit $m
\rightarrow 0$.  The mass
regulator breaks parity and even after it is sent to zero, for half-filling it induces a non-vanishing  Hall conductivity sign$(m)/4 \pi$ for each double-cone, generated by its lower filled band. In the standard case of pure $\pi$ flux the overall system preserves time reversal  and the signs in the two cones are opposite, in agreement with Nielsen-Ninomyia theorem \cite{NN}. For the holon system in consideration, however, time reversal is broken by the staggered structure of the optimal spinon configuration and by the chirality of the antiferromagnetic spin vortices. (The full spinon-holon system is nevertheless invariant under time-reversal and parity due to the opposite signs of the two Chern-Simons actions). The symmetry structure of the holon system is however preserved by time-reversal combined with  parity w.r.t. a line between the lattice sites realized by exchange of the two  N\'{e}el sublattices. As a consequence when a regulator respecting this symmetry is introduced, still denoted $m$, the contributions of the two cones to the Hall conductivity add up \cite{PRB19} to sign$(m)/2 \pi$, as in  the case of topological insulators \cite{Hal}. Taking $m>0$, in terms of effective action this corresponds to a contribution  $S_{c.s.}^{U(1)}(b)$.

In the standard free case if the upper bands are partially filled, their contribution to the Hall conductivity exactly cancel that of the lower bands. In our case, however, the situation is different, because the interaction with the slave-particle gauge field makes the holon liquid of the upper bands incompressible and with vanishing contribution to the Hall conductivity. In fact,
one should not introduce a coupling of the ``spurious'' holons in the lower bands with spinons  so that the density of ``physical'' holons coupled to spinons still correctly vanishes at $\delta = 0$; a coupling between spinons and lower bands would introduce unphysical interactions in the model. 
Hence the contribution to the Hall conductivity of the lower bands is correctly computed as above. The upper bands, however, have  additional couplings to the slave-particle gauge field, generated by spinons, and to $V^m$. In doing the computation it will be convenient to take into account the approximation of neglecting $U$ in $V^m$, by replacing  $V^m$ in the action by a $U(1)$ gauge field $v$ corresponding to the
$U(1)$ subgroup of the spin group $SU(2)$ previously selected
by choosing the directions of the spinons. Accordingly, we need also to replace the $SU(2)$ Chern-Simons  action of $V$ by $2 S_{c.s.}^{U(1)} (v)$, where the factor 2 is due to a normalization needed passing from
$SU(2)$ to its $U(1)$ subgroup. Then the interaction with $A$ and $v$, upon integration of ``physical'' fermionic holons, produce as leading term of the effective action the Chern-Simons action
  $-S_{c.s.}^{U(1)}(b +A)-S_{c.s.}^{U(1)}(v)$ .
  We now need to integrate $A$ to find both the Hall conductivity and the compressibility of the holon
system.

Since the upper bands are partially filled, the leading contribution
comes from a region near the Fermi surface. Then at $\omega =0$ in
the limit $k \rightarrow 0$ in the Coulomb gauge its polarization
bubble matrix is given by:
\begin{eqnarray}
\label{eq14}
\pi^h(\vec k)=\begin{pmatrix}
  \chi_0^h  &  -k_2 \sigma_H^h &  k_1 \sigma_H^h\\  k_2 \sigma_H^h &  \vec k^2 \chi_\perp^h & 0 \\ -k_1 \sigma_H^h & 0 &  \vec k^2 \chi_\perp^h
\end{pmatrix},
\end{eqnarray}
where $\chi_0^h,\chi_\perp^h,\sigma_H^h=-1/(2 \pi)$ are the density of
states at the Fermi energy, the diamagnetic susceptibility and the Hall
conductivity, respectively.

Since the spinons are gapped, integrating them out one obtains a Maxwell
effective action for the slave-particle gauge field $A_\mu$; the
spinon polarization bubble matrix at $\omega =0$ is then given by
the diagonal matrix
\begin{eqnarray}
\label{eq15}
\pi^s(\vec k)={\rm diag}( \chi_0^s k^2,  \vec k^2 \chi_\perp^s, \vec k^2 \chi_\perp^s),
\end{eqnarray}
with $\chi_0^s,\chi_\perp^s$ the electric and the diamagnetic
susceptibility of the spinon system.

 We now compute the polarization bubble of the
holons dressed by the spinon interaction in RPA approximation, which should give the leading contribution since the slave-particle interaction is of long-range. In the small $k$
limit it is then given by
\begin{eqnarray}
\label{eq16}
\Pi^{h}(\vec k)=  \pi^h[1+(\pi^s)^{-1}\pi^h]^{-1},
\end{eqnarray}
where the scalar component (corresponding to compressibility) is given by
\begin{eqnarray}
\label{eq17} \Pi^h_0(\vec k)=  \chi_0^s k^2,
\end{eqnarray}
and the Hall polarization bubble by
\begin{eqnarray}
\label{eq18} \sigma_H(\vec k)=\frac{\sigma_H^h k^2 \chi_0^s
\chi_\perp^s}{(\sigma_H^h)^2+ \chi_0^h (\chi_\perp^s +
\chi_\perp^h)};
\end{eqnarray}
therefore both vanish at $k=0$, implying that the Fermi liquid in the upper bands of
``physical'' holons is incompressible and does not contribute to the Hall
conductivity. Since the lower holon bands are completely filled we get as result 
that the total holon system before it is coupled to the Chern-Simons
$b$-field is incompressible and with Hall conductivity $1/(2
\pi)$.  Hence, according to the result stated in Theorem 2, after coupling with $b$ the resulting semionic holons have exclusion statistics parameter 1/2, as anticipated above. 

Since the crossover from the pseudogap PG to the strange metal SM is due to spinons, the contribution of the lower bands remain unchanged in the two "phases". In the partially filled band in SM qualitatively the situation is as discussed above, except for the value of the Fermi momenta, which is now that of the tight-binding Fermi surface.
For later purposes we finally remark that the total effective Chern-Simons action given by the original $U(1) \times SU(2)$ term of the gauged $t$-$J$ model plus the induced  action obtained integrating out the holons according to the above computation is given by   $-S_{c.s.}^{U(1)}(b)+S_{c.s.}^{U(1)}(v)$.

\section{The retarded 2D hole correlator}
In this section we sketch the computation \cite{Bert}, within some approximation, of the scaling limit of the retarded two-point correlation function of the hole reconstructed from spinons and holons in the PG "phase", near the Fermi surface. At first we neglect the contribution of the $v$ and $A$ fields, which will be re-inserted later on.

We start  noticing that in the scaling limit the contribution of the lower band of "spurious" holons is just a matrix element suppressing the spectral weight of the hole outside of the magnetic Brillouin zone. It will not be discussed anymore, for details see \cite{PAM04}.

%, also to compare it with the 1D analogue in the next Section.
To derive the contribution of the "physical" holons to the euclidean $c$-correlator in the scaling limit we use a "tomographic decomposition" \cite{FG} near the Fermi surface, a sort of dimensional reduction angle-dependent allowing to rewrite the holon correlator as a sum of contributions from 1+1 massless  Dirac fields. This procedure is an analogue of the decomposition in 1D in terms of left and right movers. More precisely let $\Vec{n}_\theta$ a unit vector from the center of the considered Fermi surface to a point on the Fermi surface at an angle $\theta$ say respect to the $x^1$ positive direction. We set $\Vec{n}_\theta \cdot \Vec{x} = x_\parallel^\theta$ and $\Vec{n}_\theta \wedge \Vec{x} = x_\perp^\theta$. Notice that $\Vec{n}_{\theta+\pi}=-\Vec{ n}_\theta$ and, accordingly,$x_{\parallel}^{\theta+\pi}=-x_{\parallel}^\theta$. Let's for simplicity omit in the following the flavor index of the Dirac fields $\psi_a$. Following ideas developed in \cite{FG, PAM04} in the scaling limit we write $\psi (x)  \exp[i \int_{\gamma_x} b] \sim \int d\theta \exp[i k_{F \theta} x_\parallel^\theta] \exp[i \int_{\gamma_x} b]  \psi_\theta (x)$, where the $\psi_\theta$ can be identified as the quasi-particle fields in the sense of Landau. Here  $k_{F \theta}$ is the Fermi momenta along the $\theta$ direction of the Fermi surface of a spin 1/2 Fermion, in spite of the fact that the holon is spinless, since, as discussed above, the semionic holons turn out to obey the Haldane statistics of parameter 1/2. The fields  $\psi_{\theta}, \psi_{\theta +\pi}$ are the analogue of the 1D chiral components $h_{R},h_{L}$ and we call them sector fields. The field $\psi_{\theta}$ can then be decomposed as a product of two fields: a fermionic $\psi_{\parallel \theta}(x^0, x_\parallel^\theta)$ and a bosonic $\psi_{\perp \theta}(x^0, x_\perp^\theta)$.  The second field has a propagator given simply by an approximate Dirac-delta function in $ x_\perp^\theta$. As in 1D one can then group together  $\psi_{\parallel \theta}$ and $\psi_{\parallel \theta +\pi}$ into a two-component massless Dirac field which we denote by $\Psi_\theta$ and we call  radial sector field. In the scaling limit  $b$ is minimally coupled only to these fields. Taking into account the renormalization of the coefficient of the Chern-Simons action discussed at the end of the previous section, the integration of the $b_0$ component produces the constraints
\begin{eqnarray}
\label{eqb}
	\frac{1}{2\pi} \epsilon_{\mu \nu} \partial_\mu b_\nu (x^0,x_\parallel^\theta,x_\perp^\theta) = \delta(x_\perp^\theta) \Psi_\theta^\dagger \Psi_\theta  (x^0,x_\parallel^\theta )
\end{eqnarray}
with $\partial_\mu= \partial/\partial x^0, \partial/\partial x_\parallel^\theta$. After choosing the gauge $b_\perp^\theta=0$ equation (\ref{eqb}) is solved by 
\begin{eqnarray}
b_\parallel^\theta (x^0,x_\parallel^\theta,x_\perp^\theta) =  \pi  {\rm sgn}(x_\perp^\theta) \Psi_\theta^\dagger \Psi_\theta (x^0,x_\parallel^\theta )
\end{eqnarray}
with the convention ${\rm sgn}(0)=0$.

Since in $\Psi_\theta$ we have assembled together the fields corresponding to angles $\theta$ and $\theta + \pi$, naively one expects that the $\theta$ parameter for the radial sector fields should range over an interval of length $\pi$. However, when we include the strings the result would depend, through a sign, on which specific interval we choose in the natural range of $2 \pi$. This situation is the analogue of the choice in 1D of the relative angle between the direction of the string and the $x^1$ axis. As we have done there we should avoid this unphysical feature and integrate the contributions of the $\Psi_\theta$ fields with $\theta$ running in the full $2 \pi$ range.

Using the Schwinger formula as in \cite{ye} and restoring physical units $x^0 \to v_f x^0$ with $v_f$ the Fermi velocity, we find for the leading contribution of the euclidean correlator of the radial sector fields:
 \begin{eqnarray}
  \langle \exp[i \int_{\gamma_x} b] \psi^*_\theta (x^0, x_\parallel^\theta)  \exp[-i \int_{\gamma_0} b] \psi_\theta(0) \rangle =	\frac{1}{ \sqrt[4]{  (v_f x^0)^2 + (x_\parallel^\theta)^2 }  }.
  \end{eqnarray}
We can see that because of the Chern-Simons coefficient $-1$ we get the well known expression of the propagators of hard-core bosons in one dimensions: the $b$ field has "bosonized" the radial sector fields. This change of statistics from the semionic one for the holon fields to the bosonic one  for the quasi-particle excitations over their ground state is reminiscent of the change of statistics appearing in the fractional quantum Hall effect, where the ground state is described in terms of fermionic electrons, but the quasi-particle excitations above it are the anyonic Laughlin vortices (see e.g. \cite{Jain}). 

Since the hole  must still be a fermion and, as discussed above, the large-scale field for the spinon is bosonic, it is up to the spin gauge field $v$ to restore the right statistics of holon quasiparticles. In fact if we add the $v$-strings and use the straight-line Gorkov approximation (see e.g. \cite{ln}) for the coupling of holons to $v$, then the action $+S_{c.s.}^{U(1)}(v)$ has precisely the right coefficient to transmute back to fermionic the exchange statistics of the quasi-particle holon fields dressed by the $v$-strings, without changing, however, the power law of the sector fields.

 We finally need to re-insert the coupling to the slave-particle gauge field $A$ and to the spinon.
 Preliminarily we remark that in the 1D systems of the radial sector fields, because of the appearance of the factor ${\rm sgn}(x_\perp^\theta)$ with ${\rm sgn}(0)=0$, all the local gauge-invariant correlation functions not involving gauge strings are unmodified by the coupling to the Chern-Simons fields. Since the leading term of the effective action of the slave-particle gauge field obtained integrating the sector fields has as coefficient polarization bubbles that are gauge-invariant without gauge strings, its scaling behaviour is the same of a fermionic system without Chern-Simons couplings.
 
 In the Coulomb gauge, the Coulomb interaction of the scalar component is screened by the finite density of holons.  The dominating term is then that of the transverse field $A^T$, with correlator given
for $\omega, |\vec k|, \omega/|\vec k| \sim 0,$ by
\begin{eqnarray}
   \langle A^T A^T \rangle (\omega,\vec k) \sim (- \chi |\vec k |^2 + i
   \kappa \frac{\omega} {|\vec q|})^{-1},
\end{eqnarray}
where $\chi \equiv \chi_\perp$ is the diamagnetic susceptibility and $\kappa$ the Landau damping.
This behavior dominates at large scales over the Maxwellian term due to the massive spinons,
destroying confinement, as anticipated \cite{IL}. Nevertheless the
attraction generated by $A^T$ in the
spinon-holon pairs is sufficient to produce near the Fermi surface a resonance with the quantum numbers of the hole. Thus in 2D a true spin-charge separation is not realized in this approach.

Unfortunately a perturbative treatment of gauge fluctuations would be insufficient to get a resonance out of a spinon and a holon, but a possible way out is to implement the  binding using an eikonal approach \cite{PAM04}. The eikonal resummation is obtained by treating first $A^T$ as an external field, using the Gorkov approximation for the holon and expanding the correlation function of the spinons in terms of
first-quantization Feynman paths. One then integrates out $A^T$ to obtain an interaction between these paths and the straight path of the Gorkov approximation, interaction
which is then treated in the eikonal approximation. Finally a Fourier transform is performed to
get the retarded correlation function.  

Plugging as typical energy scale $\mathrm{Max}(T, \omega)$, the transverse gauge propagator has a typical momentum scale: the so-called Reizer \cite{rei} momentum, $Q(T, \omega) \approx  (\kappa  \mathrm{Max}(T, \omega)/\chi)^{1/3}$.
It turns out numerically that in the PG "phase" the spatial Fourier transform
is  dominated by a nontrivial complex $|\Vec{x}|$-saddle point precisely near the inverse Reizer momentum, due to the effect of gauge fluctuations \cite{PAM04}. Below
this momentum scale, or alternatively beyond the inverse momentum scale in space,
there is an effective attraction mediated by $A^T$ between opposite charges relative
to the slave-particle $U(1)$ group, yielding in particular a hole resonance. The main effect of the complex saddle point in the retarded hole correlator is to induce a shift in the mass of spinons, $m_s \rightarrow  M_s (T, \omega)  \approx  (m_s^2 - i c  \mathrm{Max}(T, \omega)/ \chi)^{1/2}$, where $c$ is a real constant $O(1)$, which in turn adds to the chemical potential of the holon. Since in the range considered $T, \omega << m_s$,  the real part of $M_s \approx m_s$ essentially renormalizes the chemical potential and the imaginary part  introduces
a dissipation $\Gamma(T, \omega)$, proportional to $\mathrm{Max}(T, \omega)$.

A last effect that we need to consider in the PG "phase" is the attraction between the holons, and, as a consequence of the binding to spinons, between the holes due to the attraction between the attached antiferromagnetic spin vortices centered on opposite
N\'{e}el sublattices. Such attraction mathematically is generated from the term (\ref{eq10}) averaging the spinons: $\int d^2x V^m_\mu V^m_\mu \langle s^*_\alpha s_\alpha \rangle \sim \sum_{i , j} (-1)^{|i|+|j|} h^*_ih_i h^*_j h_j \Delta_{(2)}^{-1}(i-j)$, where $\Delta_{(2)}$ denotes the 2 D lattice laplacian. This attraction induces the formation of charge pairs, the spin degrees of freedom being still unpaired above a crossover spin-pairing temperature. The non-vanishing density of charge pairs is described by a $d$-wave order parameter, $\Delta_h$,  whose modulus (up to its $d$-structure) we can keep constant near the Fermi surface, but whose phase is strongly fluctuating, since the charge pairs are not condensed. In fact the field describing the phase fluctuations has a gap, which we denote by $m_\phi(T)$ decreasing with $T$, that modify the standard BCS form of self-energy near the hole  Fermi surface to
\begin{eqnarray}
\label{eq19}
\Sigma(\omega, \vec k) \approx \frac{|\Delta_h(\Vec{k},T )|^2}{\omega -\omega_h(\vec k)-m_s+ i \Gamma(\omega,T)}[1- \frac{m_\phi(T)}{\sqrt{-(\omega+ i \Gamma(\omega,T))^2+(\omega_h(\vec k)+m_s)^2+m_\phi(T)^2}}],
\end{eqnarray}
where $\Delta_h(\vec k)$ is the $d$-wave holon-pair order parameter and $\omega_h(\vec k)$ the holon dispersion \cite{MG}.
Hence finally the retarded hole correlator near the Fermi surface in the PG "phase" is given by
\begin{eqnarray}
\label{eq20}
G^R(\omega, \vec k) \approx Z(\Vec{k}) \frac{\sqrt{m_s \kappa}}{\omega -\omega_h(\vec k)- m_s+ i \Gamma(\omega,T)-\Sigma(\omega, \vec k) },
\end{eqnarray}
with $Z(\Vec{k})=1-(\cos{k_x}-\cos{k_y})/(\sqrt{2}(\cos{k_x}^2+\cos{k_y}^2))$ the direction-dependent wave-function renormalization constant arising from the mentioned matrix element due to the $\pi$-flux.

The independence of the  numerator of (\ref{eq20}) from $T$ and $\omega$ is due to the semionic nature of the holons. If the effect of the $b$ field is neglected, as in the more approximate treatment of \cite{MG}, there is a dependence in the form Max$(T, \omega)^{1/6}$, which actually disagree with experimental data in the cuprates. Let us just mention that, since the  $\pi$-flux is no more felt by physical holons, in the SM "phase" $Z(\Vec{k})$ is absent and the holon dispersion is the tight-binding one.

%%%%%%%%%%%%
\section{Comparison between 1D and 2D}
The spin-charge gauge approach to the $t$-$J$ model outlined in the previous sections allows a unified treatment of the one- and two-dimensional cases.  Its justification in terms of optimization of energy is the same in both. Furthermore the charge carriers in 1D and 2D obey an exclusion statistics with parameter 1/2, as a consequence of the no-double occupation constraint, though the origin of this property is somewhat different. In 1D it originates from the cancellation of the spinon Fermi momenta in the two chiralities due to the spin string where the constraint is used. In 2D it arises from the incompressibility of the holon fluid due to the Coulomb interaction generated by the spinons, remnant of the Gutzwiller constraint at large scales.

However, as we have seen, the final physical result of this approach are quite different in the two cases. In 1D spinon and holon are decoupled, both exhibiting semionic braid statistics, furthermore the spinon is gapless. In 2D spinon and holon are bound together at low energy, the exchange statistics of the quasi-particle bounded together are bosonic and fermionic for the spin and charge degrees of freedom, respectively.
Furthermore the spin carriers are gapped. Let us trace the origin of these differences.

First of all 
in 1D a gauge field has
no transverse component, while in 2D it does have one. The disappearance of this degree of freedom in 1D w.r.t. 2D
induces the following effects in one-dimension:
$B^T$ =0, hence there is no Hofstatder mechanism generated by  $B^m$ and the holon doesn't have a Dirac structure; $V^T$=0, hence there is no spinon mass generation, which in 2D is generated by the "magnetic" field strength of $V$; $A^T$=0, hence spinons and holons are decoupled, whereas in 2D the transverse slave-particle gauge field is able to introduce a weak binding. In fact, the other natural possible source of binding is the Coulomb interaction mediated by the scalar component of the slave-particle gauge field, but it is screened. In 2D this occurs trivially because of a Fermi surface of finite length of the holons; in 1D if one considers the CP1 non-linear sigma model describing the large-scale behaviour of spinons we have seen in equation (\ref{eq8}) that it has a $\Theta$-angle of $\pi$. Precisely at this value of $\Theta$ in 1D there are two degenerate ground states connected by a parity transformation (see e.g. Ref. \cite{AM}) as in the Heisenberg chain. Kinks interpolating between the two destroy the Coulomb attraction generated by $A_\mu$, which would be massless since the global $O(3)$ spin symmetry is unbroken, in agreement with the Mermin-Wagner theorem.
(A hedgehog gas discussed in  Ref.\cite{sac} would appear in 2D instead of the $\Theta$-term, actually inducing confinement, but it disappears when we add the holons \cite{IL} and therefore here will not be considered.)

Finally in 2D w.r.t. 1D in the considered approximation, including the interaction with the spin gauge field $v$, the braid statistics of the quasi-particle charge carriers is modified from semionic to fermionic because the finite density of semionic  holons introduce a renormalization of the effective Chern-Simons coefficient for the charge and spin gauge fields $b$ and $v$ at large scales, a phenomenon discussed also in \cite{ye15}, which has no analogue in 1D.

We close this section noticing that whereas in 1D the string approach of Weng \cite{Weng97, Weng07} is quite similar to the one reviewed here, in 2D the two approaches are instead quite different, in spite of some surviving similarities as the $\pi$-flux, the spin-vortices, not generating however a charge-pairing mechanism, the appearence of an half-filling RVB structure playing a role in some aspect analogous to the "spurious" holon bands and the recently advocated resonance nature of the hole \cite{Weng19}. In fact, the two approaches give rise to different phase diagrams of the crossovers in the $\delta$-$T$ plane, in particular in Weng's string approach there is no analogue of the pseudogap crossover line $T^*$ crossing the superconducting dome. Furthermore no fractional Haldane statistics appears in the string approach in 2D, at odds with its appearance in 1D.

%%%%%%%%%%%%%%%%%%%%%%%%%%%%%%%%%%%%%%%%%%
\section{Relevance for the cuprates}

In this section we briefly outline properties of hole-doped cuprates that found a natural explanation within the spin-charge gauge approach to the 2D $t$-$J$ model presented above, in particular due to the composite nature of the hole. Of course to describe at least semi-realistically the cuprates we need to add to the $t$-$J$ Hamiltonian at least a next-nearest-neighbour hopping term, with coefficient denoted by $t'$, but it does not modify qualitatively the general features previously discussed \cite{MG}. We also remark that, according to the observations made in Sect. 5, for the physical quantities not involving directly the hole resonance, one can consider still valid the computations made in a previous approximate treatment where the 1/2 Haldane statistics was assumed, but the effect of the $b$ field was somewhat inconsistently neglected.

We have seen that the 1/2 Haldane statistics of charge carriers is strictly linked to the presence of the charge vortices, converting spinless fermions into semions. In turn, due to the fermionic statistics of the hole, the charge vortices imply the presence of the antiferromagnetic spin vortices that provide a somewhat novel mechanism of superconductivity. In fact, in this approach superconductivity arises through a three-step mechanism.

1) At a temperature, denoted by $T_{ph}$,  charge pairing occurs as a consequence of the Kosterlitz-Thouless like attraction between antiferromagnetic spin vortices centered on holons, corresponding to empty sites in the $t$-$J$ model, on opposite  N\'{e}el sublattices. From the explicit expression (\ref{eq19})(\ref {eq20}) of the hole Green function, since the value of $m_{\phi}$ decreases with
$T$, one finds that lowering the temperature there is a gradual reduction of the spectral weight on the Fermi surface
at small frequency as we move away from the diagonals of the Brillouin zone, due
to the $d$-wave structure of $\Delta_h(\vec k)$. Simultaneously at larger frequencies we have the
formation and increase of two peaks of intensity precursors of the excitations in
the superconducting phase \cite{MG}. Thus the main effect of charge-pairing is the generation of a phenomenology of Fermi arcs near the diagonals, qualitatively consistent with ARPES
experiments in underdoped cuprates (see e.g. \cite{norman})

2) At a temperature denoted by $T_{ps}$, lower than $T_{ph}$, using the holon-pairs as sources of attraction, the slave-particle gauge attraction between holon and spinon 
induces the formation of short-range spin-singlet (RVB) spinon pairs. Hence, below $T_{ps}$ there is a finite density of hole pairs.

3) Finally, at an even
lower temperature,  $T_c$, the hole pairs
become coherent and a $d$-wave hole condensate (in BCS approximation) appears,
leading to superconductivity.

Hence, in the $\delta$-$T$ phase diagram of the present approach to the 2D  $t$-$J$ model, besides the antiferromagnetic transition not considered here, there is a phase transition, the superconducting one, and three crossovers corresponding to charge-pairing, spin pairing and transition from a "Dirac" Fermi surface to a tight-binding one due to an effective "melting" of the $\pi$-flux. A comparison of the emerging structure of the phase diagram with experimental data in hole-doped cuprates is given in Fig.1.
 The theoretical computations whose result are presented in Fig.1 are
performed  with experimental inputs for one doping to fix parameters that are then used consistently
in all the calculations in each "phase". $T_{ph}$ and $T_{ps}$ are evaluated solving gap equations in Mean-Field approximation, see \cite{PRB11}.

\begin{figure}
\begin{center}
\includegraphics[width=6.5cm]{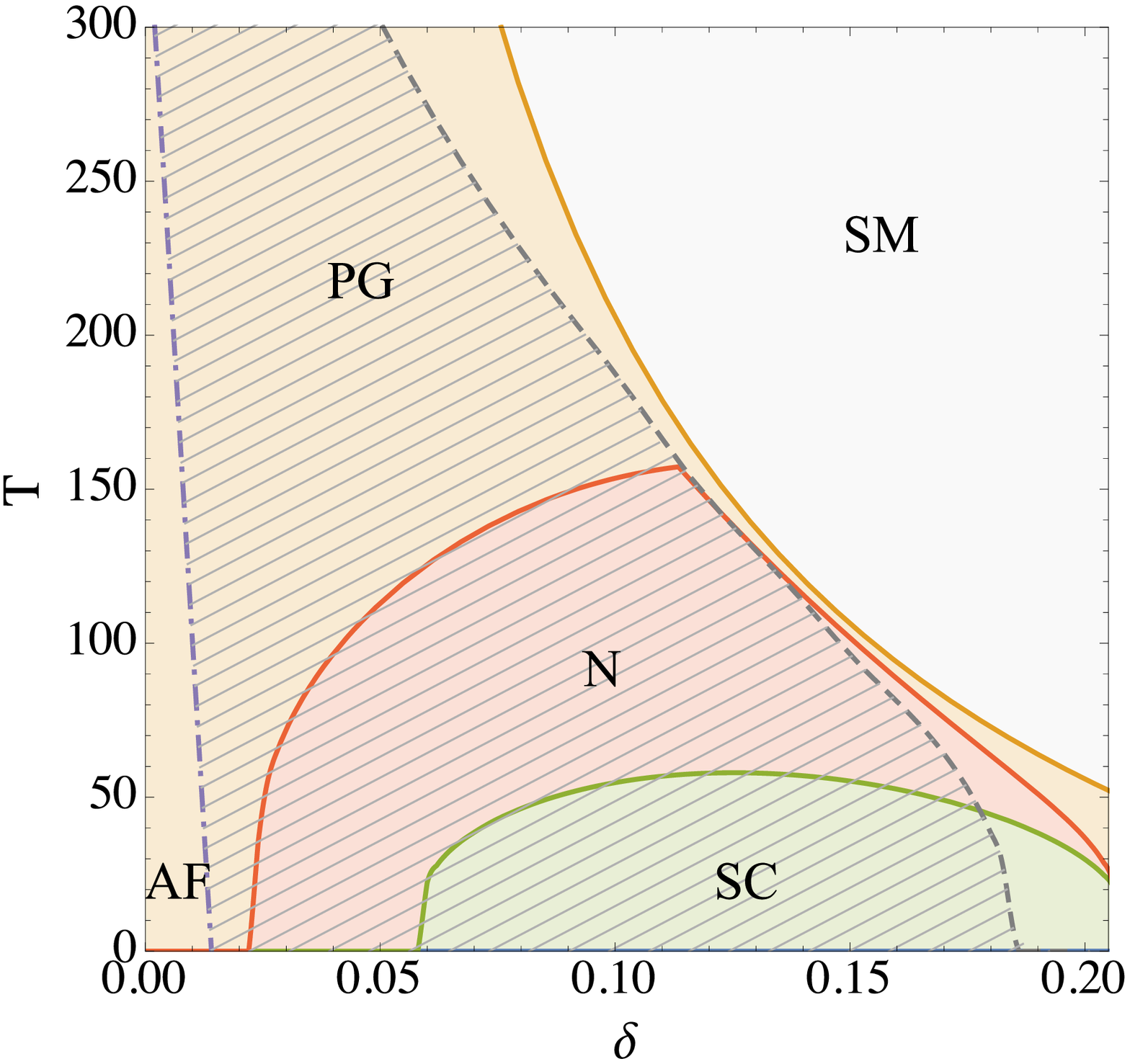} \\a\\
\includegraphics[width= 6.5cm]{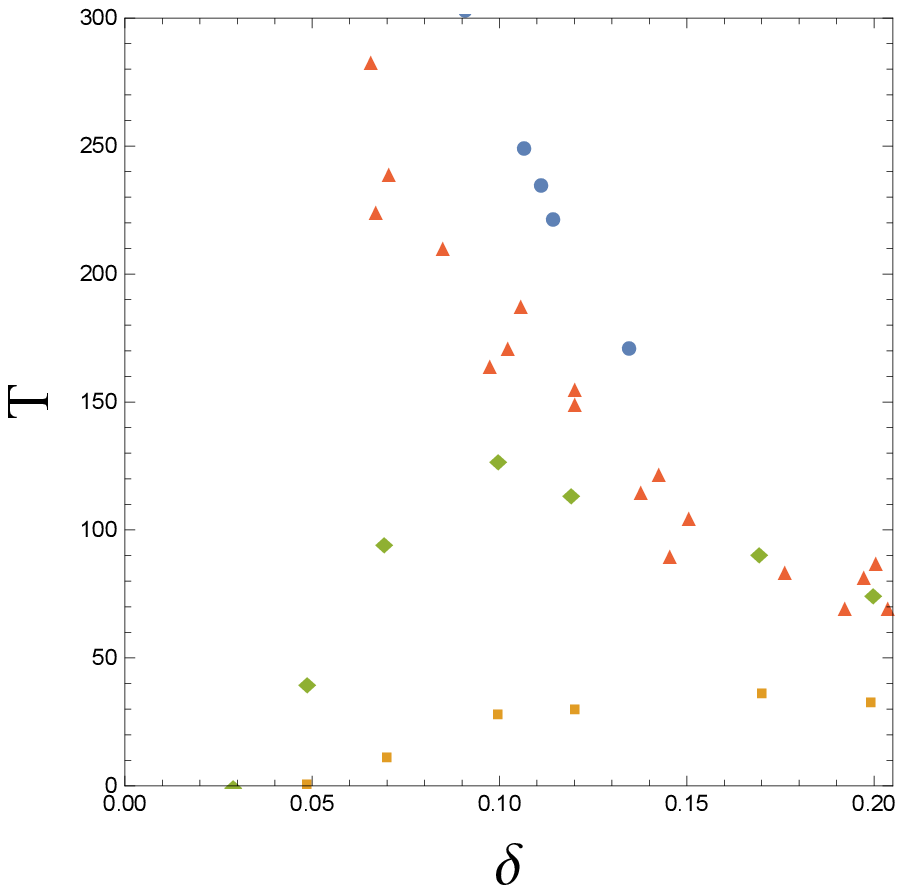} \\b
  \caption{(\textbf{a}) Theoretically derived phase diagram from Ref. \cite{MB}: holon pairing temperature $T_{ph}$ (yellow line) , spinon pairing temperature $T_{ps}$ (red line) , $T_c$ (green line). The N\`eel temperature (dot-dashed line) is qualitative from experiments not derived theoretically. (\textbf{b}) Experimental data for $T_c$ (yellow squares) and onset of Nernst signal (green diamonds) in LSCO from Ref.\cite{liwan}, ``low pseudogap'' (red triangles) in LSCO from Ref.\cite{ho} and ``high pseudogap'' (blue circles) in YBCO from Ref.\cite{ba}.}%
 % \label{fig1.2}
\end{center}
\end{figure}

The comparison between the experimental data and the theoretical curves suggests the identification of $T_{ph}$ with the "upper pseudogap temperature" corresponding for example to the deviation from linearity of the in-plane resistivity (see e.g. \cite{ando}). This crossover line does not cross the superconducting dome and the interpretation in terms of charge-pairing shares some analogy with Uemura's approach \cite{Ue}, at odds instead with the RVB interpretation (see e.g. \cite{lee}) in terms of spin-pairing. Similarly $T_{ps}$ appears to be identifiable with the onset of magnetic-field induced
diamagnetic  and Nerst signals due to vortices \cite{liwan}, appearing in presece of  "preformed hole pairs", in agreement with many other interpretations. The line of "melting" of the $\pi$-flux is seen to correspond to the "lower pseudogap temperature" marked for example by the inflection point in the in-plane resistivity \cite{ando}. It crosses the superconducting dome as advocated  in many approaches based on the existence of a critical point (see e.g.\cite{Tal}). The only experimentally observed crossover that is completely missed in this approach is the onset of charge-density waves peaked around $\delta =1/8$ (see e.g. \cite{Tail}); we conjecture that this is due to a contribution of the oxygen orbitals in the cuprates not taken into account by the $t$-$t'$-$J$ model.

Besides the structure of the phase diagram there is another general feature of data of the cuprates that found a natural explanation within our approach \cite{MB}: a non Fermi-liquid universality \cite{uni} of many suitably normalized experimental curves, indicating an independence of details of the Fermi surface. In fact, if the quantity we compute in the spin-charge gauge approach depends essentially on spinons and holons by themselves and  not on the hole as a resonance and if the spinon contribution is the dominating one,  universality is a natural consequence of the fact that the spinon propagator  does not depend on details of the Fermi surface. The first of the above requirements holds if the Ioffe-Larkin composition rule \cite{IL} is valid. Both the requirements are satisfied for example in  the case of in-plane resistivity and superfluid density in the PG "phase". In fact the theoretically derived universality curves are in a quite satisfactory also quantitative agreement with the corresponding experimental data, as shown in Fig.2, with the experimentally observed "3/2" critical exponent for the superfluid density \cite{hardy} well reproduced, due to the 3DXY nature of the spinon-triggered superconductivity transition \cite{MB1}.

\begin{figure}
\begin{center}
\includegraphics[width=6.5cm]{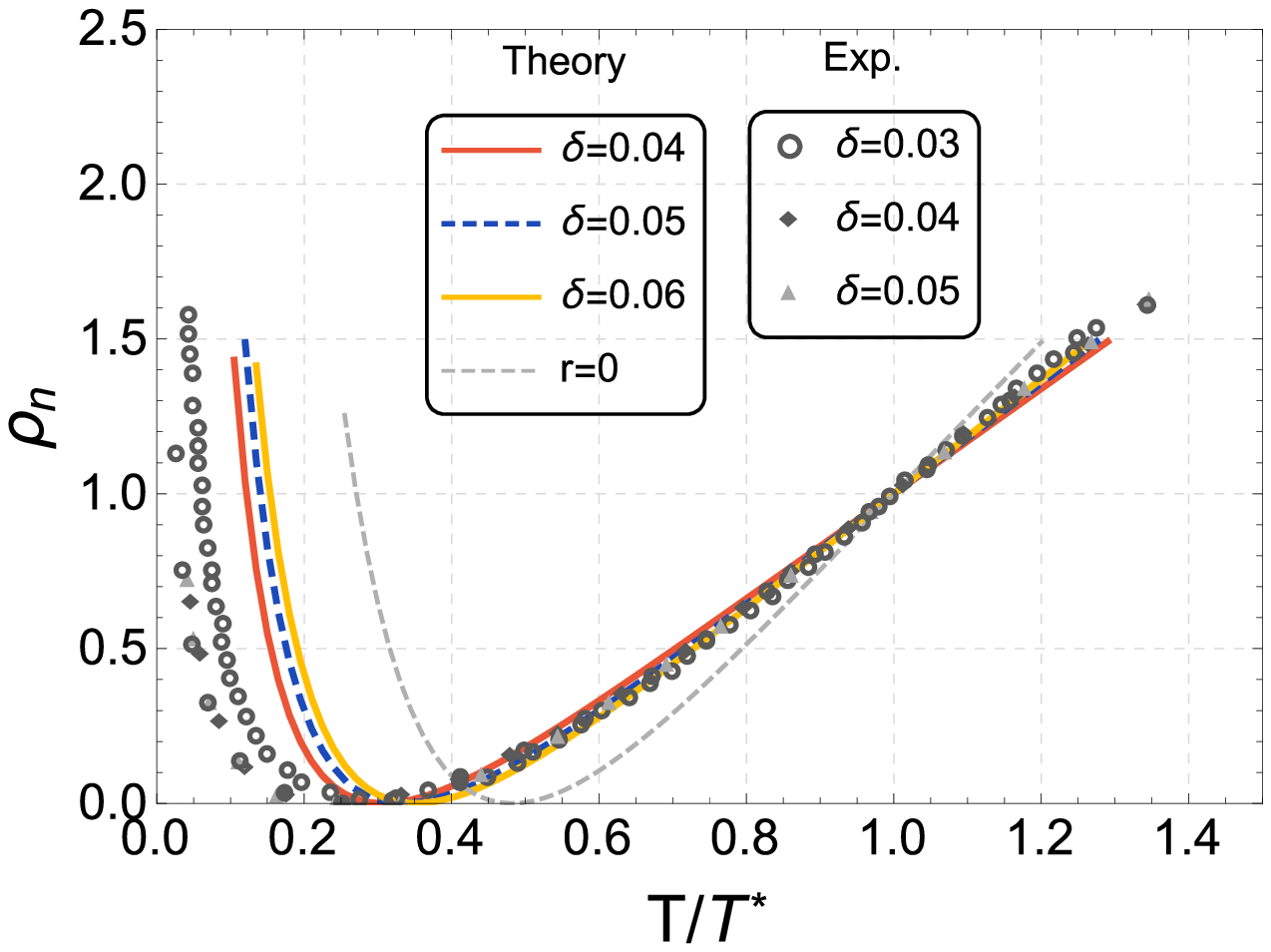}\\a\\
\includegraphics[width=6.5cm]{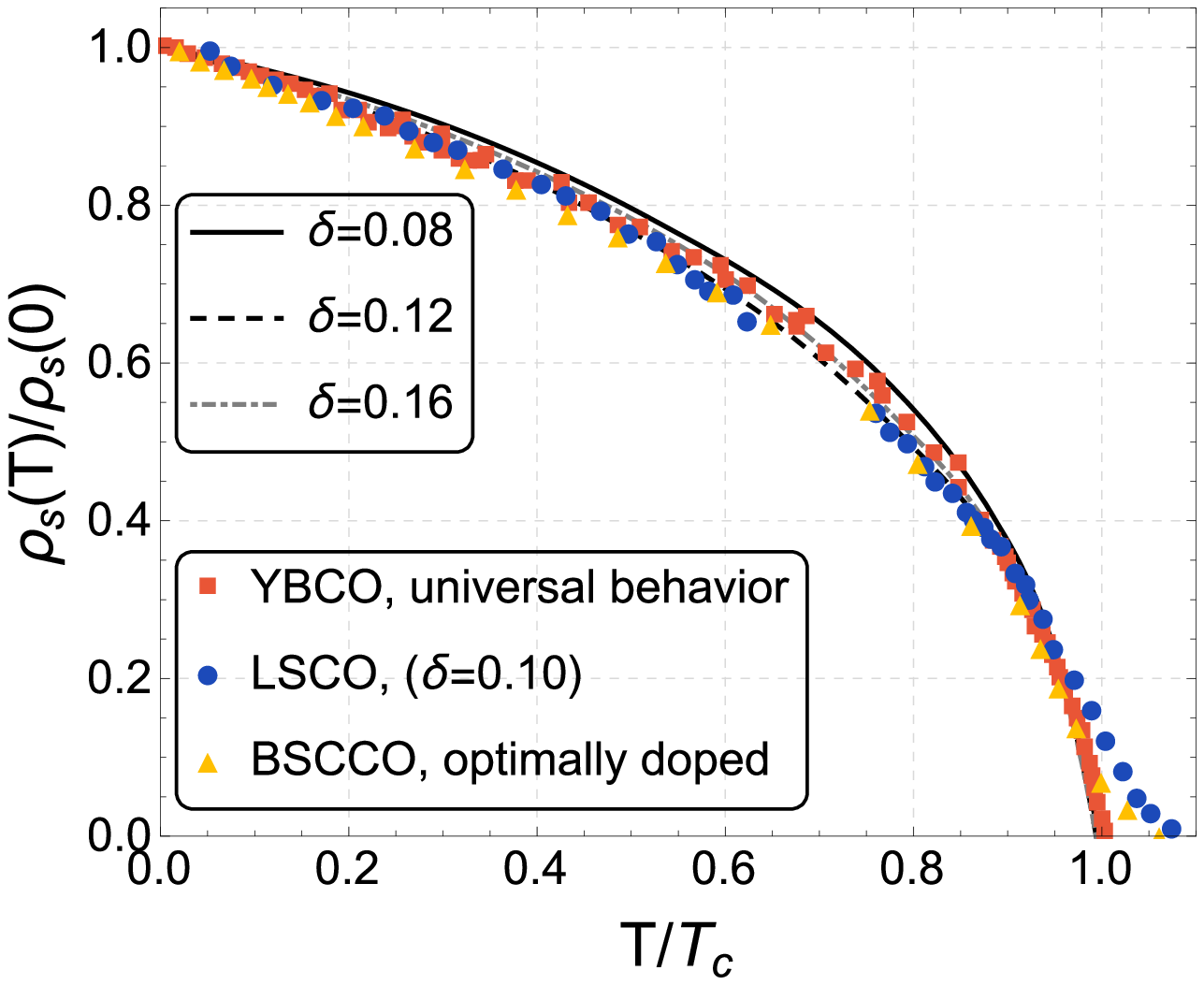}\\b 
  \caption{(\textbf{a}) The normalized in-plane resistivity $\rho_{n}$ theoretically calculated, including a holon contribution with relative weight $r$ w.r.t. the spinon contribution, from Ref. \cite{MB}, compared with experimental data from Ref. \cite{ando}. The discrepancy at low $T$ might be due to the missing account of holon and spinon pair formation in the above treatment. (\textbf{b}) The normalized superfluid density $\rho_s$ from Ref. \cite{MB}, compared with experimental data for underdoped (LSCO, YBCO) and optimally doped (BSCCO, YBCO) samples, from Refs. \cite{hardy}\cite{ja,pa}. }%
%  \label{fig1.2}
\end{center}
\end{figure}

Among other experimental features of cuprates that find a natural explanation are crossovers appearing in transport quantities in the PG "phase" and the emergence of Fermi arc phenomenology, briefly discussed in the previous section. Concerning the first feature,  examples are the metal-insulator crossover of the in-plane resistivity \cite{PAM01,MB} and the peak in the $^{63}(1/T_1T)$  spin-lattice relaxation rate \cite{mdsz}.
Both are interpreted as due to the competition between the spinon gap dominating at lower $T$ and the dissipation growing with $T$ introduced by the couping of the spinons to the slave-particle gauge field.

We are presently analyzing \cite{Bert} within this approach another quite remarkable feature of the cuprates: the coexistence of the non-Fermi liquid character of many quantities with a  Fermi-liquid-like behaviour of other quantities, in particular in the SM phase, as for example for the uniform magnetic susceptibility. We  conjecture that this occurs because in these last cases the dominant contribution comes from the hole resonance and not from the spinons.

The above overlook to the comparison between the theoretical scheme sketched in this review and experimental data in the cuprates shows that, although exhibiting an admittedly  complicated structure, the approach based on the 1/2 exclusion statistics for the charge carriers of the 2D $t$-$J$ model is able to explain naturally many unusual features of the hole-doped cuprates.

%\vfill
%%%%%%%%%%%%%%%%%%%%%%%%%%%%%%%%%%%%%%%%%%
{\bf Acknowlegments.} I gratefully acknowledge J\"{u}rg Fr\"{o}hlich whose original intuition was crucial for the origin of this journey attempting to understand the physics of the cuprates, Su Zhao-Bin and Yu Lu, for the pleasure of a collaboration since the beginning of this journey, Ye Fei who later brilliantly joined us and the many other collaborators involved in the project, in particular Dai Jian-Hui, Lorenzo De Leo, Giuliano Orso, Michele Gambaccini, Alberto Ambrosetti, Giacomo Bighin and Tommaso Bertolini.

%%%%%%%%%%%%%%%%%%%%%%%%%%%%%%%%%%%%%%%%%%

%\vfill


%=====================================
% References, variant A: internal bibliography
%=====================================
%\reftitle{References}

\end{document}